\documentclass[[journal,12pt,onecolumn]{IEEEtran}
\IEEEoverridecommandlockouts
\usepackage{url}
\usepackage{subfig}
\usepackage{cite}
\usepackage{amsmath,amssymb,amsfonts}
\usepackage{algorithmic}
\usepackage{graphicx}
\usepackage{textcomp}
\usepackage{adjustbox}
\usepackage{graphicx}
\usepackage{xcolor}
\usepackage[textsize=tiny,textwidth=1.4cm]{todonotes}
\usepackage{mathtools,lipsum, nccmath,cuted}
\usepackage{siunitx}
\usepackage{textcomp}

\usepackage{comment}
\usepackage{makecell}
\usepackage{algorithm}
\usepackage{algorithmic}
\usepackage{gensymb}
\usepackage{ulem}
\usepackage{colortbl}
\usepackage{enumitem}
\definecolor{Gray}{gray}{0.925}
\usepackage[textsize=tiny]{todonotes}

\graphicspath{ {figures/} }
\def\BibTeX{{\rm B\kern-.05em{\sc i\kern-.025em b}\kern-.08em
    T\kern-.1667em\lower.7ex\hbox{E}\kern-.125emX}}
    
\usepackage{pgfplots}
\usepgfplotslibrary{colorbrewer}
\pgfplotsset{compat = 1.14, cycle list/Set1-8} 
\usetikzlibrary{pgfplots.statistics, pgfplots.colorbrewer,pgfplots.dateplot,
        shadows,
        matrix} 

\usepackage{pgfplotstable}
\usepackage{filecontents}
\usepackage{graphicx}
\pgfplotsset{compat=1.14}

\definecolor{blueLine}{RGB}{57,106,177}
\definecolor{blueFill}{RGB}{114,147,203}
\definecolor{redLine}{RGB}{204,37,41}
\definecolor{greenline}{RGB}{0,250,0}
\definecolor{blackLine}{RGB}{0,0,0}
\definecolor{goldLine}{RGB}{160,82,45}

\usepackage[font=footnotesize]{caption}
\begin{document}

\title{Malaria detection using Deep Convolution Neural Network}

\author{
    \IEEEauthorblockN{Sumit Kumar\IEEEauthorrefmark{1}, Sarthak Kapoor\IEEEauthorrefmark{2}
    , Harsh Vardhan\IEEEauthorrefmark{3}, Sneha Priya\IEEEauthorrefmark{4}, Ayush Kumar\IEEEauthorrefmark{5}\\}
    \IEEEauthorblockA{\IEEEauthorrefmark{1}Georgia State University, \IEEEauthorrefmark{2}Amazon,\IEEEauthorrefmark{3}Vanderbilt University,\IEEEauthorrefmark{4}ICFAI University, \IEEEauthorrefmark{5}Amity University}
}

\maketitle

\section{Abstract}
{According to the WHO's \textit{2023 World Malaria Report}, malaria cases rose to 249 million—an increase of 2 million from the previous year. Global efforts to combat malaria appear to have stalled, primarily due to a decline in international funding. Malaria, transmitted through bites from infected female mosquitoes, is prevalent in 91 countries, with nearly 90\% of infections and deaths occurring in sub-Saharan Africa. The disease claimed 435,000 lives last year, most of whom were children under the age of five in Africa.
Artificial intelligence has brought transformative advances to malaria diagnosis in parts of Africa. The Malaria Cell Image Dataset, sourced from the official NIH website, was developed to support microscopists in low-resource settings and to enhance diagnostic precision via AI-powered algorithms that detect and segment red blood cells.
This work demonstrates that even a simple 2-layer convolutional neural network (CNN) can achieve state-of-the-art accuracy in malaria detection, setting a new benchmark for AI applications in this field. A middleware component is also developed to enable deployment of the model on mobile platforms. Additionally, an Out-of-Distribution (OOD) detector, trained on the same dataset, is employed to identify anomalous inputs and flag them for manual review, thereby reducing false positives. The trained CNN model achieves 95.4\% accuracy on test data for classifying malaria versus non-malaria cases. The RRRCF-based OOD detection mechanism performs with 90\% efficiency. Together, the integrated system yields an overall malaria prediction accuracy of 99.5\%.  
}

\section{Introduction}
In engineering, model-based design\cite{neema2019web}, model-based control\cite{brosilow2002techniques}, and model-based optimization\cite{vardhan2019modeling} have long served as foundational tools for building intelligent, explainable, and reliable systems. With the emergence of large-scale data and computational power, data-driven approaches, particularly deep convolutional neural networks (CNNs), have revolutionized a range of application domains. CNNs have demonstrated impressive results in image classification tasks such as ImageNet \cite{krizhevsky2017imagenet}, and have been successfully applied to problems in engineering design\cite{vardhan2022data}, autonomous driving\cite{bojarski2016end, vardhan2021rare}, radiological diagnostics\cite{gross1990neural}, and human genome analysis\cite{sundaram2018predicting}, driving the development of state-of-the-art control and prediction systems\cite{vardhan2021rare}.
CNNs have achieved strong performance in image-based medical diagnostics due to their ability to extract hierarchical spatial features. The core component, the convolutional layer, consists of learnable filters that are trained to extract hierarchical features from input images. The breakthrough of AlexNet in the 2012 ImageNet competition \cite{krizhevsky2017imagenet} marked a turning point in deep learning. Successive architectures such as GoogLeNet \cite{ballester2016performance} further pushed the boundaries, achieving a classification error rate as low as 0.06656 and approaching human-level accuracy.
This work focuses on leveraging CNNs to build an end-to-end system for automated malaria detection from cell images, specifically distinguishing parasitized from uninfected red blood cells. Our experimental results show that a lightweight CNN model, when carefully designed and tuned, can outperform larger models previously used in this domain. This offers a compelling advantage in real-world deployments where computational resources are limited.
To enable practical adoption, we also require further tooling and develop a middleware framework for mobile deployment of the trained CNN model. This includes integration with platforms such as TensorFlow Lite and ONNX for real-time inference on smartphones and tablets, which are often the primary computing devices in low-resource or remote areas. The middleware handles pre-processing, inference, and post-processing to ensure an efficient and user-friendly experience for healthcare workers in the field.
Furthermore, a critical safety component of safety critical prediction system is the incorporation of out-of-distribution (OOD) detection to flag anomalous, corrupted, or irrelevant inputs. CNNs are known to make overconfident predictions on data outside the training distribution, which can lead to critical errors in medical applications.
In summary, this work presents a complete pipeline for accurate, efficient, and deployable malaria diagnosis using CNNs, emphasizing model compactness, deployment feasibility on mobile devices, and the trustworthiness of predictions through OOD detection. Such a system holds significant promise for improving early malaria diagnosis in underserved regions.

\section{Problem Statement}

Despite significant advances in machine learning for medical imaging, malaria continues to pose a major health threat in resource-constrained regions due to lack of rapid, accurate, and scalable diagnostic tools. Manual examination of blood smear slides by trained microscopists is time-consuming and error-prone, particularly in remote areas with limited access to healthcare infrastructure. The integration of deep convolutional neural networks (CNNs) into mobile-based diagnostic workflows presents a promising solution, but critical challenges remain in model accuracy, deployment feasibility, and reliability under real-world conditions. This research work aims to address three interlinked problems in the development of an end-to-end malaria diagnostic system powered by CNNs:

\subsection{CNN-Based Malaria Detection}
The first objective is to develop a robust CNN model that accurately classifies red blood cells as parasitized or uninfected from microscopic blood smear images. The model needs to be trained on publicly available and/or custom-labeled datasets, with architectural choices evaluated for sensitivity, specificity, and inference efficiency. Data augmentation, class imbalance handling, and regularization techniques is also employed to improve generalization. Sub-problem is to train a model and to optimize CNN performance for malaria detection given class imbalance, label noise, and varying image quality?

\subsection{Middleware Development for Mobile Deployment}
The second objective is to build an efficient middleware layer that enables seamless integration of the trained CNN model into mobile platforms, such as Android or iOS. The middleware will handle model inference using lightweight libraries (e.g., TensorFlow Lite, ONNX, CoreML), pre- and post-processing pipelines (image normalization, patching, visualization), and resource management for real-time usage in low-power environments. Sub-problem that we are addressing here to ensure high-speed, low-latency CNN inference on edge devices with limited memory, processing power, and inconsistent connectivity?

\subsection{Out-of-Distribution (OOD) Detection for Diagnostic Safety}
The third objective addresses the critical need for \textit{input reliability verification}. Deep learning models are known to produce high-confidence predictions even on inputs vastly different from their training distribution (e.g., corrupted, low-resolution, or irrelevant images). To prevent erroneous diagnoses, an OOD detection module will be developed to flag anomalous or invalid inputs before classification.  Sub-problem of interest is how to detect and reject out-of-distribution samples in real time on mobile devices, without significantly increasing computational cost?

The goal of this work is to develop a trustworthy, mobile-compatible, and field-deployable malaria diagnostic system that combines the accuracy of CNNs, the accessibility of mobile health platforms, and the robustness of anomaly-aware inference for real-world clinical utility. We are addressing all three above mentioned problem in this work.

\section{Approach}
\begin{figure}[h!]
        \centering
        \includegraphics[width=0.9\textwidth]{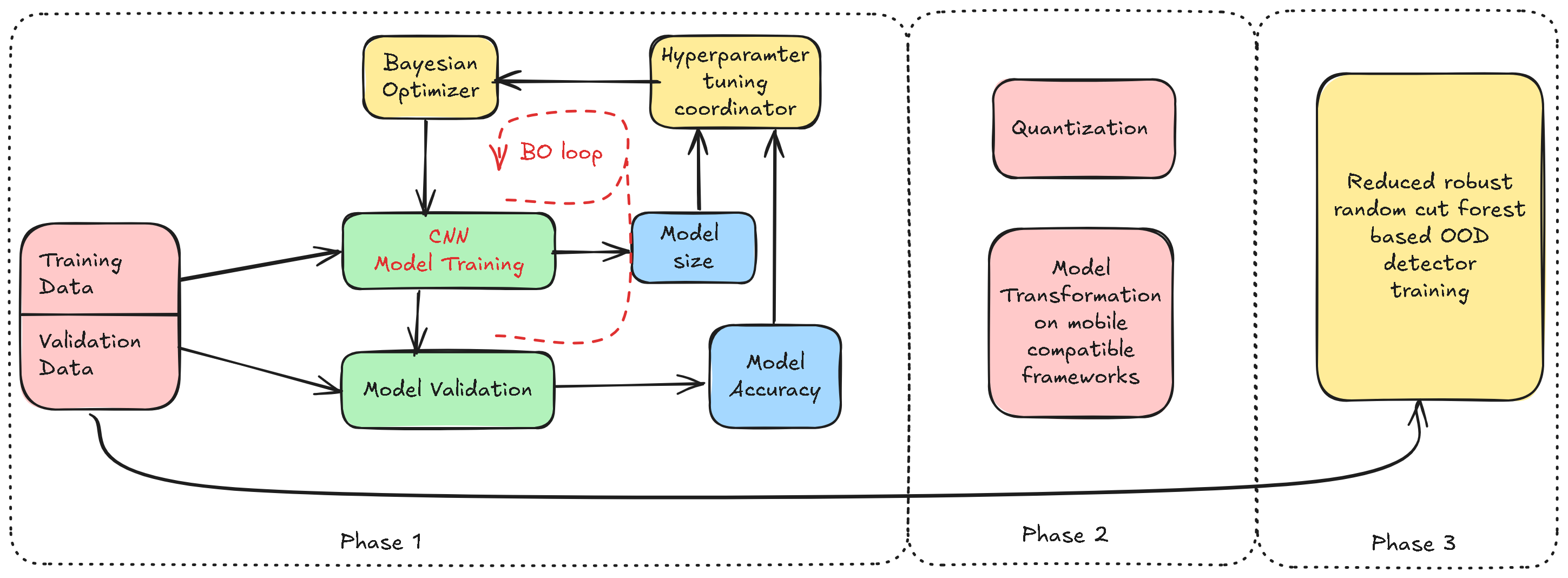}
        \caption{The three phased approach to the problem. CNN Model training phase trains the model for classification task. After \textit{CNN Model training}, the \textit{model validation} predicts the accuracy of validation data, that BO loop kicks in for hyperparamter tuning optimization phase, that optimize the CNN architecture to minimize the model size and maximise the model accuracy.}
        \label{fig:final_arch}
    \end{figure}

We segeregate our approach in three sections or phases (refer fig \ref{fig:final_arch}). In first phase (Phase 1), the primary objective in this phase is to construct a lightweight, resource-efficient CNN model capable of distinguishing parasitized and uninfected red blood cells (RBCs) from microscopic images, optimized for eventual deployment on mobile platforms.

For data preparation, we used the publicly available NIH Malaria Dataset, containing 27,558 Giemsa-stained thin blood smear images (balanced across parasitized and uninfected classes). The images undergo pre-processing by first being resized to 64 × 64 pixels, which reduces computational load while preserving key spatial features necessary for discrimination. They are then normalized to have zero mean and unit variance across the RGB channels to ensure consistent input distribution. To improve the model's ability to generalize and to mitigate overfitting, data augmentation techniques such as rotations, flips, and brightness adjustments are applied and generated additional 10,000 data set.
\begin{figure}[h!]
        \centering
        \includegraphics[width=0.4\textwidth]{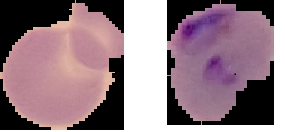}
        \caption{Input sample data : Image of cell (left : Not infected cell) ; (right : Infected cell)}
        \label{fig:sample_data}
    \end{figure}

The model architecture is carefully crafted following a thorough hyperparameter tuning process using Bayesian optimization (called \textit{BO loop}) with the dual optimization objective of maximize the validation accuracy and minimize size of neural network. 

To effectively navigate the hyperparameter subspace, we utilize Bayesian Optimization with an Upper Confidence Bound (UCB) acquisition function. This approach balances exploration of uncertain regions with exploitation of known high-performing configurations. The UCB is formally expressed as:
\begin{equation}
\text{UCB}(x) = \mu(x) + \kappa \cdot \sigma(x)
\end{equation}
where \( \mu(x) \) and \( \sigma(x) \) denote the predicted mean and standard deviation of the objective function at point \( x \), and \( \kappa \) controls the trade-off between exploration and exploitation. This formulation ensures that the optimization process remains sample-efficient while converging toward high-performing model architectures.

The goal is to train a lightweight yet effective neural network design consisting of convolutional blocks and fully connected hidden layers. 
Each convolution layer consist of layer of convolution neurons,max pooling layer, and batch normalization layer. 
The number of such convolutional layers and the number of convolution neurons in these layers are treated as parameters in hyperparameter tuning.The CNN layer is followed by the flattened layer, and the number of flattened layers and the number of neurons in a layer are also parameters that would be optimized in hyperparameter tuning loop. It concludes a total four dimensions of hyperparameter subspace trained using BO-UCB (Bayesian optimization-Upper Confidence Bound). The design ensures a judicious balance between model complexity and computational efficiency, making it well-suited for deployment in resource-constrained environments without dedicated GPU support.

In each iteration of \textit{CNN Model Training}, the model is trained by minimizing cross-entropy loss using the  Adaptive Moment Estimation (Adam) \cite{kingma2014adam}, with the learning rate finely adjusted and exponentially reduced within the range of $[1e-4, 1e-3]$ to balance convergence speed and stability. To mitigate overfitting and promote efficient learning, early stopping is used along with a learning rate decay schedule that reduces the learning rate as training progresses. One iteration of the hyperparameter tuned training process involves 50 epochs, using batch sizes ranging from 32 to 64, providing a robust balance between gradient estimation accuracy and computational efficiency. The loss function deployed is categorical\_crossentropy. 


We ran 50 iterations of BO loop as well. 

The final optimized model after training completion of Phase1, It begins with an input image of size 64 x 64 x 3, which is passed through a convolutional layer containing $32$ filters of size 3 × 3 with a stride of $1$, producing a feature map of dimensions 62 x 62 x 32. This is then downsampled via a 2 x 2 max-pooling layer\cite{murray2014generalized}, reducing the spatial size to 31 × 31 × 32, thereby lowering computational demand while retaining essential spatial features. To further enhance training efficiency and stability, a batch normalization layer is applied across all feature channels, followed by a dropout layer\cite{baldi2013understanding} with a rate of $0.2$ to mitigate overfitting through random neuron deactivation \cite{murray2014generalized}. This entire sequence—convolution, pooling, normalization, and dropout—is repeated in a second block with similar structural adjustments.  The output of the final block is then flattened and passed through a dense layer with softmax activation to generate class probabilities, distinguishing between parasitized and uninfected cells. 
After the two complete convolution layers, the feature size has reduced enough and most of the feature is extracted so that we can now connect it with the feed-forward network. The $14\times14\times32$ feature matrix is flattened, whose size becomes 6272. This flattened feature is fed to a feed-forward network of size 512, along with batch normalization and dropout. The feed-forward network is further added to another layer of 256 neurons with batch normalization and dropout. Finally, the output layer is connected with 2 neurons, and the activation function in the output layer is softmax. The activation of the rest of the layers is 'Rectified Linear unit(RELU)\cite{agarap2018deep}. The cost function for the error measurement is used as categorial\_crossentropy, and the optimizer is  Adaptive Moment Estimation (Adam) \cite{kingma2014adam}, which combines ideas from both RMSProp and Momentum. It computes adaptive learning rates for each parameter.   

The model's performance is assessed using standard classification metrics called classification accuracy, to capture both overall and class-specific predictive quality.  Additionally, practical deployment considerations are addressed by evaluating the model size in megabytes and inference latency on a mobile CPU, ensuring the model meets the constraints and responsiveness requirements of resource-limited environments.

\begin{figure}[h!]
        \centering
        \includegraphics[width=0.8\textwidth]{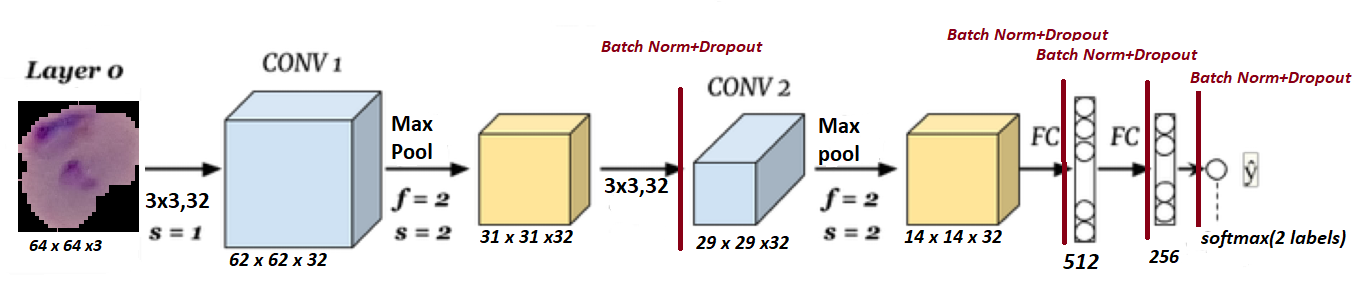}
        \caption{Final CNN model after Phase 1 of training.}
        \label{fig:sample_data}
    \end{figure}

\section{ Cross device Domain adaptation and Mobile based  deployment for light-weight CNN}
In the second phase of our work (called Phase 2 in figure \ref{fig:final_arch}), cross-device domain adaptation and mobile-based deployment are addressed to ensure the robustness and practicality of our lightweight CNN model for real-world use. Although the model is trained on standardized image data under controlled settings, deployment scenarios—particularly on mobile or edge devices—introduce domain shifts due to variations in camera quality, lighting conditions, and environmental noise. To mitigate this, domain adaptation techniques can be integrated to align feature distributions between the source (training) and target (deployment) domains, enhancing generalization across devices. Furthermore, to meet the constraints of mobile hardware, our model is explicitly designed to be lightweight, with minimal parameter count and reduced computational complexity. To directly deploy on mobile platform,  the trained model is exported to TensorFlow Lite and ONNX formats for Android-based deployment, with optional INT8 quantization to further reduce memory footprint and inference time.  

\section{Outlier monitor training}
In phase 3, the detection of outliers in high-dimensional image sequences is addressed via a Reduced Robust Random Cut Forest (RRRCF), an efficient variant of the Robust Random Cut Forest tailored to large datasets \cite{vardhan2022reduced}. Initially, RRRCF constructs a compact ensemble of cut‑trees that preserves the geometric structure of the training data while significantly reducing memory overhead. During offline training i.e. Phase 3 in figure \ref{fig:final_arch}, samples are selectively included—or discarded—based on their displacement value, which quantifies the increase in tree complexity when the sample is inserted; only samples that exceed a threshold contribute to the forest. 

To quantify this increase, we define a dispersion metric that captures the perturbation a new sample introduces to the existing forest structure. Mathematically, the dispersion value for a given input \( x \) is computed as:
\begin{equation}
\text{Disp}(x) = \mathcal{C}(T \cup \{x\}) - \mathcal{C}(T)
\end{equation}
where \( \mathcal{C}(T) \) represents the complexity of the RRRCF forest \( T \), measured in terms of entropy-based structural statistics. A higher dispersion score indicates a significant deviation from the learned distribution, signaling that the input is likely out-of-distribution. This quantitative measure enables reliable anomaly detection while maintaining computational efficiency.

In inference, each incoming image frame is temporarily inserted into the prebuilt forest: the change in model complexity (DispValue) is measured, and if it exceeds the learned threshold, the frame is labeled as out-of-distribution and flagged as anomalous. The method's suitability for image streams is demonstrated using simulated precipitation scenarios (e.g., rain intensity), where RRRCF reliably distinguishes between in‑distribution data images frames and those which are out of distribution. We selected a cut forest with tree size of 20, and criteria to measure the quality of a split  is entropy using the Shannon information gain formulation. This approach enables robust, interpretable OOD detection in streaming image data without extensive hyperparameter tuning, making it well-suited for real‑time deployment. For more details on reduced robust random cut forest, please refer this paper \cite{vardhan2022reduced, guha2016robust}. 

\begin{figure}[h!]
        \centering
        \includegraphics[width=0.9\textwidth]{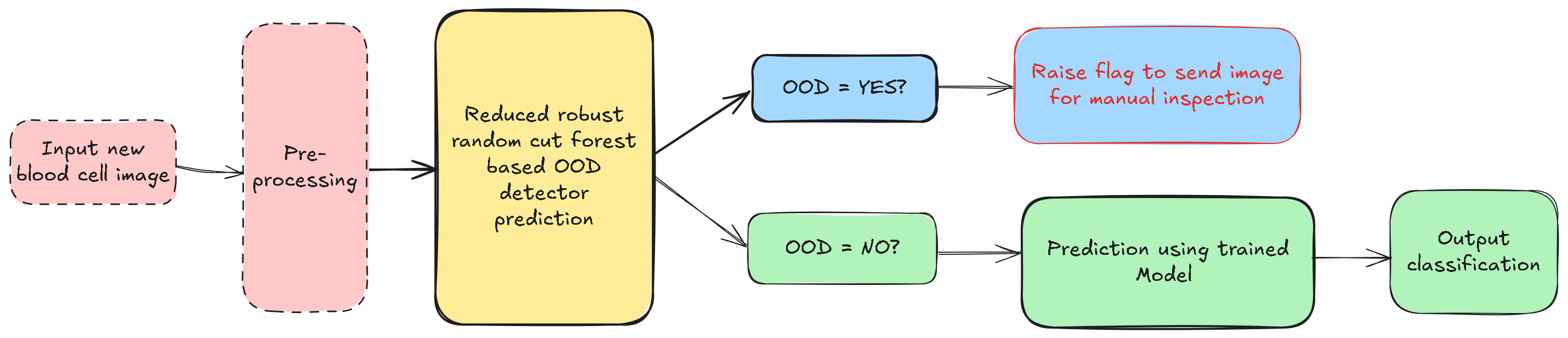}
        \caption{The runtime execution order}
        \label{fig:runtime}
    \end{figure}

\section{Results}

\begin{figure}[h!]
        \centering
        \includegraphics[width=0.45\textwidth]{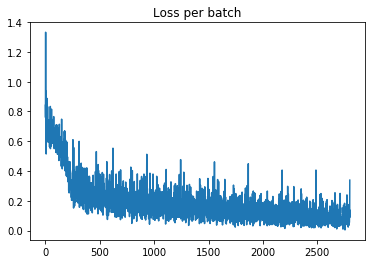}
        \caption{Plot of cost with respect to batch in one iteration of CNN model training.}
        \label{fig:cost}
    \end{figure}

The training and validation error are calculated simultaneously as the training progress. The number of epochs is set to 50, while a callback function is written to invoke early stopping when the model's validation accuracy reaches 95\% of accuracy. This is done to avoid overfitting (generalization error). 
Once the training is done (which stops when validation accuracy reached 95\%), then the testing error is calculated on the trained model. The testing accuracy achieved around 95.4\%. The plot of the cost function is plotted with respect to batch progression (refer figure \ref{fig:cost}). The result of batch training is the cost becomes noisy and fluctuates, but the overall trend of progression is toward optimal value.

After training, the runtime execution order is setup (shown in figure \ref{fig:runtime}) that starts with pre-processing of input image. The processed image is fed to RRCF based detector that detects and flag if the input data is qualified for OOD or not.  If OOD flag is raised, the data is sent to the manual inspection in laboratory. If OOD is false, the pre-processed cell image data is passed to the trained CNN model that predicts the output i.e. the classification of whether the cell is having malaria parasite or not. 

\subsection*{Integrated System Accuracy Calculation}

Our CNN model achieves a test accuracy of 95.4\%. To enhance clinical safety, we incorporate an Out-of-Distribution (OOD) detector that flags anomalous or uncertain inputs with 90\% detection efficiency. This mechanism ensures that most misclassified or noisy samples are identified before reaching the classification stage.

The combined effective accuracy of the system, accounting for both correct classifications and correctly flagged anomalies, is computed as:

\begin{equation}
\text{Effective Accuracy} = 1 - \left[(1 - \text{CNN Accuracy}) \times (1 - \text{OOD Recall})\right]
\end{equation}
\begin{equation}
= 1 - (0.046 \times 0.10) = 0.9954
\end{equation}

Thus, the integrated system achieves \textbf{99.54\% effective prediction accuracy}, meaning it either produces a correct result or flags a sample for manual review when uncertain.


\section{Related work}
Machine learning is being used in control\cite{vardhan2022reduced,duan2016benchmarking,wang2012machine,vardhan2023search,wu2019machine,duriez2017machine} to model\cite{neema2019design,audet2000surrogate} and prediction of complex systems \cite{vardhan2021machine,volk2020biosystems,vardhan2022deepal,mazurenko2019machine,vardhan2023constrained,moosavi2020role,vardhan2022deep}. 
There are various works in using machine learning in Malaria detection \cite{gopakumar2018convolutional,gourisaria2020deep} used deep CNN for Malaria detection but their model is more complex than ours. \cite{gopakumar2018convolutional} used stacking-based approach for automated quantitative detection of Plasmodium falciparum malaria from blood smear. For the detection, a custom designed convolutional neural network (CNN) operating on focus stack of images is used. The cell counting problem is addressed as the segmentation problem and we propose a 2-level segmentation strategy. Use of CNN operating on focus stack for the detection of malaria improved the detection accuracy (both in terms of sensitivity [97.06\%] and specificity [98.50\%]) but also favored the processing on cell patches and avoided the need for hand-engineered features. \cite{poostchi2018image} used machine learning for detection of Malaria. 
 In 2016, Uganda's Ministry of Health found that the disease is the leading cause of death in the country - accounting for 27 per cent of deaths.
Mortality rates are particularly high in rural areas, where the lack of doctors and nurses is acute. Nursing assistants are often taught to read slides instead, but inadequate training can lead to misdiagnosis. Due to lack of availability of lab technicians in the region lead some to process four times as many as recommended number of screening in a day, while it is recommended that each technician should process no more than 25 slides each day. 
There are so many patients who may require malaria and TB tests, and overworking day and night. The AI lab(\cite{uganda}, at Makerere University, has developed a way to diagnose the blood samples using a cell phone.
The program learns to create its own criteria based on a set of images that have been presented to it previously. It learns to recognize the common features of the infections. The smartphone clamped in place over one microscope eyepiece brings to light a detailed image of the blood sample below - each malaria parasite circled in red by artificially intelligent software. 
With this AI backed technology, pathogens are counted and mapped out quickly, ready to be confirmed by a health worker. Diagnosis times could be slashed from 30 minutes to as little as two minutes.

\section{Conclusions}
In this work, we presented a lightweight yet highly effective 2-layer deep convolutional neural network for malaria diagnosis and detection. Our 2-layer CNN achieves 95.4\% accuracy using minimal preprocessing, and integrates OOD detection to improve clinical safety. With mobile deployment support, it offers a scalable solution for malaria screening in resource-limited settings. Additionally, the integration of an out-of-distribution detection mechanism enhances the reliability of the system by flagging uncertain cases for expert review. This study establishes a strong baseline for low-complexity, AI-driven diagnostic tools and demonstrates the potential for scalable deployment, including on mobile devices, to aid frontline healthcare efforts in malaria-endemic regions.

\bibliographystyle{IEEEtran}
\bibliography{bibliography}

\end{document}